\documentclass[12pt,preprint]{aastex}
\begin{document}
%% updated to ApJ version

\title{Thermal and Nonthermal X-ray Emission from the Forward Shock in
Tycho's Supernova Remnant}
\author{Una Hwang (1,2), Anne Decourchelle (3), Stephen S. Holt (4), and Robert Petre (1)}
\affil{(1) NASA Goddard Space Flight Center, Greenbelt, MD 20771 \\
(2) Department of Astronomy, University of Maryland, College Park, MD
  20742 \\
(3) Service d'Astrophysique, CEA Saclay, 91191 Gif-sur-Yvette Cedex, France \\
(4) F. W. Olin College of Engineering, Needham, MA 02492}

\begin{abstract}

We present Chandra X-ray images of Tycho's supernova remnant that
delineate its outer shock as a thin, smooth rim along the straight
northeastern edge and most of the circular western half.  The images
also show that the Si and S ejecta are highly clumpy, and have reached
near the forward shock at numerous locations.  Most of the X-ray
spectra that we examine along the rim show evidence of line emission
from Si and S ejecta, while the continuum is well-represented by
either a thermal or nonthermal model.  If the continuum is assumed to
be thermal, the electron temperatures at the rim are all similar at
about 2 keV, while the ionization ages are very low, because of the
overall weakness of the line emission.  These electron temperatures
are substantially below those expected for equilibration of the
electron and ion temperatures, assuming shock velocities inferred from
radio and X-ray expansion measurements; the electron to mean
temperature ratios are $\lesssim0.1-0.2$, indicating that
collisionless heating of the electrons at the shock is modest.  The
nonthermal contribution to these spectra may be important, but cannot
be strongly constrained by these data.  It could account for as much
as half of the flux in the 4-6 keV energy range, based on an
extrapolation of the hard X-ray spectrum above 10 keV.

\end{abstract}

\section{Introduction}

As the bright remnant of an historically observed supernova in our
Galaxy, Tycho's supernova remnant (SNR 1572) has been extensively
studied, but the X-ray spectrum associated with its forward shock has
not been directly measured until now.  By necessity, most X-ray
spectral studies have focused on the spatially integrated spectrum,
which is dominated at energies below a few keV by the ejecta.  A faint
outer shelf of emission was identified in X-ray images from the
Einstein Observatory, however, and is attributed to material behind
the forward shock (Seward, Gorenstein, \& Tucker 1983)---an
association that is further supported by the excellent correspondence
of the X-ray boundary with the sharp outer boundary seen in radio
images (Dickel et al. 1991).

Imaging studies with the ROSAT X-ray Observatory establish that the
expansion of the remnant varies both with azimuthal angle and radius
(Hughes 2000).  The faster expansion observed at the outer boundary
corresponds to an average forward shock velocity of 4600 $\pm$ 400
($D$/2.3 kpc) km/s, where $D$ is the distance in kpc.  The expansion of
the radio boundary is also observed to vary with azimuthal angle
(Strom, Goss, \& Shaver 1982, Reynoso et al. 1997).  Strong
deceleration in the east is caused by the remnant's interaction with
dense H gas (Reynoso et al. 1999).  Although the outer boundaries of
the X-ray and radio emission show excellent correspondence, the
average expansions measured at these wavelengths are inconsistent: the
expansion rate $m$, defined such that the time evolution of the
remnant radius is $r \sim t^m$, is 0.471 $\pm$ 0.028 in the radio
(Reynoso et al. 1997), and 0.71 $\pm$ 0.06 for the outer radii in
X-rays (Hughes 2000).  This discrepancy is unresolved, but appears to
be a common pattern in young remnants (such as Cas A, Koralesky et
al. 1998, Vink et al. 1998, and Kepler's SNR, Dickel et
al. 1988, Hughes 1999).

In the case of Tycho's SNR, the forward shock velocity can be measured
in yet another way.  The optical emission is almost exclusively H
Balmer emission from nonradiative shocks propagating into partially
neutral gas (Chevalier \& Raymond 1978, Chevalier,
Kirshner, \& Raymond 1980, Kirshner, Winkler, \& Chevalier 1987).  The
neutral H atoms are not heated by the shock, and can be collisionally
excited before being ionized; these excited atoms produce narrow
Balmer lines consistent with their low temperatures.  Slow H atoms can
also undergo charge exchange reactions with fast protons that have
already been heated behind the shock; these fast H atoms contribute a
broad component to the line profile.  The optical emission from
Tycho's SNR is generally faint, and is detected only in the eastern
and northern regions.  The most prominent feature is on the eastern
side of the remnant (knot {\it g} in the compilation of Kamper \& van
den Bergh 1978).  Smith et al. (1991) and Ghavamian et al. (2001)
infer shock velocities through knot {\it g} of $\sim$ 2000 km/s
independent of the distance to the remnant.  From modelling the line
emission, while accounting for the effects of both the electron-ion
temperature equilibration and Ly $\alpha$ scattering, Ghavamian et al.
constrain the electron to proton temperature ratio to be
$\lesssim$0.10, implying electron temperatures $\lesssim$0.8 keV in
the knot.

X-ray spectral studies of Tycho's SNR generally infer a relatively
hard spectral component, presumed to be associated with the forward
shock.  The total spectrum from the Advanced Satellite for Cosmology
and Astrophysics (ASCA) X-ray Observatory suggests a forward shock
component with a temperature of roughly 4 keV that accounts for some
30\% of the X-ray flux between 0.5 to 10 keV (Hwang, Hughes, \& Petre
1998).  This temperature is well below the mean equilibrium
temperature behind a 4600 km/s shock.  Hughes (2000) points out that
this implies either a low efficiency for electron heating, so that the
electron temperature is well below the mean, or nonlinear particle
acceleration, which could result in significantly higher compressions
and lower temperatures than for test-particle shocks of the same
velocity (Decourchelle, Ellison \& Ballet 2000).  Furthermore, X-ray
emission has been observed from the remnant at energies up to 25 keV
(Pravdo \& Smith 1979, Fink et al. 1994, Petre et al. 1999), and this
emission has also been attributed to material behind the forward
shock.  This interpretation requires a heating mechanism that can
rapidly heat the electrons to sufficiently high temperatures (e.g.,
Cargill \& Papadopoulos 1988).  The hard X-ray emission has
alternatively been suggested to come from a nonthermal population of
highly energetic electrons that has been accelerated at the shock
(Aharonian et al.  2001).

X-ray instruments have lacked the capability to isolate the spectrum
of the forward shock thus far.  Recent XMM-Newton observations have
now provided the first truly spatially resolved spectra of Tycho's SNR
on angular scales of several arcseconds (Decourchelle et al. 2001).
Even higher spatial resolution (less than 0.5$''$ FWHM) is provided by
the Chandra X-ray Observatory.  In this paper, we present images and
spectra of selected portions of the forward shock of Tycho's SNR using
the Advanced CCD Imaging Spectrometer (ACIS) on the Chandra
Observatory.

\section{Images}

Tycho's SNR was observed for 48.9 ks on 20-21 September 2000 with the
ACIS spectroscopic array.  The count rate was steady, and there were
no background flares.  The remnant has a size slightly larger than the
$8' \times 8'$ field of view of a single ACIS CCD.  The northern and
eastern parts of the remnant were positioned on the back-illuminated
CCD S3, the primary spectroscopic chip, while the western region of
the remnant was imaged on the neighbouring front-illuminated CCD S2.
This placement of the source on the detector was originally chosen to
optimize observation of the ejecta knots in the east.  The
southernmost region of the remnant fell outside the field of view of
the detector.

The broadband 0.5-10 keV image of Tycho's SNR obtained by ACIS is
shown in Figure 1(a).  Also shown in Figure 1 are energy band images
covering (b) the Si He $\alpha$ blend near 1.86 keV, with the
underlying continuum subtracted, and (c) the 4-6 keV continuum
band. In forming the X-ray images, the energy selection was performed
after correcting for the spatial variation of the gain across the
detector, as summarized in Table 1.  For the Si image, the continuum
underlying the line emission was determined from the shape of the
local spectrum, and removed.  The rather sparse 4$-$6 keV continuum
image has been smoothed.  Panel (d) shows the 22 cm radio image of
Dickel et al. (1991), convolved with a beam of 1.45$'' \times 1.38''$,
and scaled to match the X-ray images in size.  ``Equivalent width''
images of the kind presented by Hwang et al. (2000) are shown in the
three panels of Figure 2 for (a) the Fe L emission (transitions to n=2
levels) near 1 keV, (b) Si He $\alpha$ blend, and (c) the Fe K blend
(transitions to n=1 levels) near 6.5 keV.  These images show the ratio
of the line emission to the underlying continuum after smoothing.

\subsection{Outer Shock}

The Chandra broadband image shows that the outer edge of the X-ray
emission in the northeast and west is strikingly traced by a thin,
smooth rim.  The rim is circular in the west, but there is a straight
section in the northeast, and the rim is absent altogether in the
east, where the remnant is interacting with dense H gas (Reynoso et
al. 1999).  The rim is also present at radio wavelengths, where it
extends to the southern edge of the remnant that was not imaged by
this Chandra observation (Figure 1d, Dickel et al. 1991, Reynoso et
al. 1997).  Overall, the radio rim corresponds closely with the X-ray
rim, but the two do not track each other particularly well in
brightness, as illustrated in Figure 3, which compares the surface
brightness along the rim in the X-ray continuum image with that in the
radio image (also see Dickel et al. 1991).

X-ray radial profiles for two azimuthal sectors in the northwest and
southwest are shown in Figure 4 for both the broadband and 4-6 keV
continuum images.  The rim marks the location of the remnant's forward
shock, and is highlighted in the 4-6 keV X-ray continuum image.  This
energy band suppresses the ejecta contribution because it is virtually
free of the emission lines that arise mostly from the ejecta and
dominate the broadband spectrum; the thick bright shell in the
broadband image disappears in the 4-6 keV continuum image.  The
full-width-half-maximum radial extent of the arcs is seen to be only
about 4-4.5$''$, compared to the remnant radius of roughly 4$'$.  The
inside half of the rim is thinner in the continuum image, consistent
with the emergence of thermal line emission at the inside edge behind
the forward shock.

There are also a number of bright ``knots'' in the hard X-ray
continuum image, most notably near the eastern edge and in the
southwest.  These knots have previously been noted in the lower
spatial resolution 4-6 keV image obtained by XMM-Newton (Decourchelle
et al. 2001).  Radio emitting knots in the eastern part of the remnant
are not at the same location as the X-ray continuum knots, but are
slightly northward.  Indeed, the radio knots appear to be better
correlated with the Si emission, as noted by Decourchelle et
al. (2001).  The dense cloud of H gas discussed by Reynoso et
al. (1999, their ``eastern knot'') is offset by about $1'$ to the
north of the eastern bright X-ray continuum knot.

\subsection{Ejecta Distribution}

The ejecta in Tycho's SNR are visible only from their X-ray emission.
The optical emission comes entirely from the forward shock, and the
infrared emission is also associated with the forward shock (Douvion
et al. 2001), while the radio emission is due to electrons accelerated
at the shock rather than to ions.  The general distribution of the
ejecta is evident in the broadband X-ray image (Figure 1a), which is
dominated by the Si line emission, and is seen even more clearly in
the Si line image of Figure 1(b). Overall, the Si emission is
distributed fairly uniformly on large scales, and is very clumpy in
appearance, but not compact or knotty like the Si ejecta emission from
Cas A (Hughes et al. 2000, Hwang et al. 2000).

To calibrate the observed Si line-to-continuum ratios in Figure 2b
with the Si abundances, we used the ACIS detector response to compute
the count rates expected from solar-abundance thermal models with
various temperatures and ionization ages.  We find that if the
line-to-continuum ratio exceeds 3-4, the Si abundance must be enhanced
above the solar value for temperatures above 0.7 keV, regardless of
the ionization age.  Lower values of the ratio could be consistent
with enhanced abundances if the temperature is below 0.7 keV,
depending on the ionization age.  In the brightest Si clouds, the
ratios are high enough that the element abundances must be well above
solar, as expected.  The finger-like projections that reach the
forward shock in the west have lower values of the line-to-continuum
ratio, but are shown to be associated with ejecta by the spectral
analysis presented in the following section.  The two large, bright,
Si- and Fe- rich ejecta knots that bulge beyond the forward shock in
the east were known prior to Chandra and XMM (Vancura et al. 1995,
Hwang \& Gotthelf 1997); it is now evident that Si ejecta reach the
forward shock elsewhere in the remnant, though without distorting the
remnant boundary.  It is plausible that many of the Si clumps seen in
the interior are similar to the finger-like features at the edges, but
are seen face on.

Most of the Si emission is clumpy on angular scales greater than
$\sim5''$, but the continuum-subtracted Si line image (Figure 1b)
shows low level emission throughout the remnant.  This emission may be
associated with a faint, smooth component of Si ejecta (see Wang \&
Chevalier 2001).  To estimate the contribution of such a component, we
smoothed the Si line image with an adjustable beam containing at least
25 counts, subtracted the off-source background, and determined the
lowest intensity level inside the remnant.  Assuming that the faint
emission is distributed uniformly over the remnant, we estimate that
it makes up about 25\% of the Si line photons.  It is possible that
this faint Si emission will appear clumpy when viewed at higher
statistical significance, but clumps with a significant density
enhancement should be readily visible since the X-ray luminosity
scales as the square of the density.  The surface brightness contrast
of the faintest visually identified clumps in the image is a factor of
two.

In comparison to the Si images, the Fe L and K line images appear to
be somewhat less clumpy, but they are also much more sparse.
Moreover, it is difficult to calibrate the Fe L image, as the numerous
Fe L lines are blended with each other and with lines of Ne and O, so
that the true continuum cannot be determined without detailed spectral
modelling.  Except for the prominent Fe-rich knot in the east, we
cannot ascertain from the images alone whether Fe L emission is
present at the forward shock.  The brightest Fe K emission is
certainly associated with ejecta, as the line-to-continuum ratio is
generally high, and any ratio greater than 2 indicates enhanced
abundances for any temperature or ionization age.  The Fe K emission
is radially interior to Fe L and Si, as has already been noted
(Decourchelle et al. 2001, Hwang \& Gotthelf 1997); as suggested by
Decourchelle et al., this is likely to be the signature of the
temperature and ionization structure behind the reverse shock.
Detailed discussion of the ejecta spectra will be presented elsewhere.

\section{Spectra of the Forward Shock}

We study the spectrum of the forward shock along the outer rim in the
southwest, northwest, and northeast, for the regions indicated in
Figure 5.  We avoid the eastern region, where the rim is not clearly
defined, and where the remnant is interacting with dense H gas.  The
spectral regions are typically $5''$ wide, corresponding to 0.05 pc at
a distance of 2.3 kpc.  With the spectra from adjacent regions behind
the rim, we sample angular distances up to about 10$''$ behind the
shock front.  The size of the spectral regions and the counts in each
spectrum are given in Table 2.  With one exception, each spectrum has
at least $\sim$2000 counts.  We bin the spectra to have at least 25
counts per pulse height channel, and fit the energy range from
0.5 keV up to 10 keV.

Using the tools in CIAOv2.2, we have customized the spectral
calibration files for each spectrum.  Since the gain and resolution
vary with position on the detector, we use appropriately weighted
detector response functions.  Spectra were taken from a single CCD
chip, with the exception of region 5, which is mostly from S3, but
includes a 15\% contribution from S2.  Although the spectral
calibration of S2 is not as complete as for S3, the results obtained
for spectra taken from S2 are consistent with those from S3.  We also
combined the nearly featureless spectra from regions 1 and 2 in the
southwest (on S2), and regions 4 and 5 in the northwest (on S2 and S3)
to improve the signal-to-noise-ratio.  Background spectra are taken
from source-free regions on the same CCD chip (or a weighted
combination from both chips, if appropriate).  We fitted the spectra
with both nonthermal and thermal models, as described in the following
sections.

It has recently come to light that ACIS suffers from a time-dependent
absorption due to the build-up of contaminants on the detector (see
http://cxc.harvard.edu).  The effects of this absorption increase with
time and can be included in either the response function or the
spectral model.  For our observation, taken only 424 days after
launch, the effect of this absorption on the fitted spectra is
relatively small.  When we include its effect in the spectral model,
we find that the interstellar absorbing column density is the only
parameter that is significantly affected, being reduced by about
5$\times 10^{20}$ cm$^{-2}$ from the values in Tables 3-6.

\subsection{Models}

For fitting thermal models, we primarily use single temperature,
single ionization age, nonequilibrium ionization models (XSPEC v11.0,
Borkowski, Lyerly, \& Reynolds 2001).  The ionization age $n_e t$ is
defined as the product of electron density and time since
shock-heating, and parameterizes the progress toward ionization
equilibrium of gas that is suddenly heated to a high temperature.
Because we take very narrow regions behind the shock front, this
simple model is an excellent approximation for the spectra at the rim.
In the cases where the fitted ionization age is very low because of
the absence of line emission, we find that a simple NEI model is
essentially indistinguishable from models that approximate the
emission behind a segment of a planar shock wave using a range of
ionization ages ($pshock$), or from a simple bremsstrahlung continuum.
Detailed comparisons are shown in Table 3 for such a case---the
featureless spectrum from the northwest rim segments 4 and 5.

For fitting nonthermal models, we favor the synchrotron radiation
cutoff model ($srcut$; see Reynolds \& Keohane 1999, Dyer et al. 2001)
over simple power-law models.  The $srcut$ model folds the single
particle synchrotron emissivity with a power-law electron energy
distribution having an exponential cutoff.  It is the most appropriate
readily-available model for describing the synchrotron spectrum with
the fewest assumptions.  It takes as parameters the radio spectral
index, the frequency at which the spectrum rolls off (to a factor of 6
below the extrapolation of the radio spectrum, Reynolds 1998), and the
radio flux at 1 GHz as the normalization.  Reynolds \& Keohane use
this model with the spatially integrated X-ray spectrum of Tycho's SNR
and the observed radio brightness and spectral index to determine the
energy at which the photon spectrum must roll off if the model is not
to exceed the observed X-ray spectrum.  The electron energy is then
determined from the rolloff frequency by taking the magnetic field to
be a fiducial value of 10 $\mu$G.  Reynolds \& Keohane estimate the
rolloff frequency to be $8.8 \times 10^{16}$ Hz on average
(corresponding to an energy of 0.4 keV), to set the upper limit on the
cutoff energy of the electron spectrum at 40 TeV.

In our use of the $srcut$ model, we take the appropriate slope of the
radio spectrum for each region of the remnant from Katz-Stone et
al. (2000).  The normalization of the model is taken to be the radio
flux at 1 GHz, and the rolloff frequency is freely fitted.  To obtain
the radio flux for each spectral region, we use the 1.36 GHz (22 cm)
image of Dickel et al. (1991).  This 1983 image is of lower angular
resolution than the Chandra image, but all the spectral regions we use
are much larger than the radio beam size.  We first scale the radio
image to match the Chandra image, and then extract the radio fluxes
from the same spatial regions used for the X-ray spectral analysis.
To convert the flux at 1.36 GHz to the value at 1 GHz, we simply took
the product of the fraction of the total 1.36 GHz flux in each
spectral region with the total 1 GHz flux of 56 Jy (Green 2001).  An
indication of the uncertainty in the radio flux is given by the
difference between the 1.36 GHz flux in our image and that determined
by extrapolating a 1 GHz flux of 56 Jy using the average radio
spectral index of 0.50 (Katz-Stone et al. 2000); this difference is
about 15\%.

Comparisons of thermal and nonthermal model fits for the northwest rim
are also shown in Table 3.  The $srcut$ model gives a somewhat better
fit than a simple power-law.  Also shown is a hybrid containing both
thermal and nonthermal components, to be discussed in more detail in
the following section.  The northwest region of the remnant is notable
for having the highest value measured for the radio spectral index, at
0.72$\pm$0.14 (Katz-Stone et al. 2000), but it is evident from our
spectral fits that the higher-than average spectral index determined
in the radio is not compatible with the radio flux together with the
X-ray spectrum in the context of the $srcut$ model.  Even allowing for
$\sim$15\% uncertainty in the radio flux normalization, an acceptable
fit requires the radio spectral index to be below the low end of its
error range, at a value closer to the overall average value found for
the remnant.  This does not seem implausible, given the difficulties
inherent in measuring the radio spectral index.

\subsection{Results for Rims}

For the complete set of spectra at and behind the rim, we present the
results for the thermal NEI models in Table 4 and Figure 6.  The
spectra in the northwest and southwest sections are pristine and show
very little evidence for any line emission.  The other rim spectra do
show weak line emission, especially from Si and S, but generally not
from Fe L; most of the spectra taken from behind the rim also show the
emergence of line emission.  We use a single NEI component with solar
abundances to represent the forward shock, unless the poor quality of
the fit warrants the addition of a second component (we choose a
criterion of $\chi^2/dof \gtrsim 1.3$).  Since line emission is
generally responsible for the poorer fits with single component
models, this second component represents ejecta.  The forward shock
temperatures obtained are very similar around the entire rim, at about
2 keV, and the ionization age is uniformly low, with upper limits in
the range of a few times $10^8$ cm$^{-3}$ s.

For cases like region 3 in the west, the line emission seen in the
spectrum is clearly from Si and S ejecta, as is corroborated by the
Si equivalent width image (Figure 2b).  The line emission can be
modelled with a NEI component containing only the intermediate mass
elements Si and S in their solar abundance ratios (with the exception
of the region behind rim 3 which requires enhanced S), and generally
without significant Fe or Ni.  This may be justified by the
observation that the Si ejecta are, on average, more abundant at large
radii than Fe ejecta (see Hwang \& Gotthelf 1997).  The fits are not
sensitive to the temperature assumed for the ejecta, as a large range
of temperatures will give satisfactory results if the ionization age
is adjusted.  We therefore warn that the ionization ages determined
for the ejecta in Tables 4-6 should not be taken too seriously.  We
used an ejecta temperature of 0.9 keV, following the results for the
globally averaged spectra (Hwang \& Gotthelf 1997).  In a few
instances, the fits are substantially improved with the addition of Fe
and Ni in the ejecta, generally with abundances well below the solar
ratio relative to Si and S (see Table 4).  In other cases, such as
regions 6 or 7, the line emission may not necessarily be from ejecta,
as lines of O, Ne, and Fe are present in addition to those of Si and
S.  This tends to be true of the regions in the north or northwest,
which are also distinguished by the presence of optical emission and
by distortions of the remnant boundary.

The fitted ionization ages behind the northern rim are higher than
those at the rim---a result one would expect since the interstellar
gas behind the rim was shocked at an earlier time.  For the northwest
and southwest regions (1, 2, 4, and 5), however, the fitted ionization
ages behind the rim remain low, with upper limits consistent with
those at the rim itself.  This much smaller (i.e., undetectable in our
data) change in the ionization age is qualitatively consistent with a
lower density for the interstellar gas in the northwest and southwest
than in the north.  In turn, this is broadly consistent with the
interaction of the remnant with denser gas to the north and east that
is indicated by the radio observations and the presence of optical
emission.  The typical radio shock velocity is about 3000-4000 km/s
all along the western edge of the remnant.  For a velocity of 3500
km/s, a distance of 5$''$ behind the shock would imply an increase in
the ionization age $n_et$ by about $6\times 10^8$ cm$^{-3}$ s for an
ambient density of 0.3 cm$^{-3}$ for Tycho's SNR (e.g. Seward et
al. 1983). Upper limits for the ionization age of $2-3 \times 10^8$
cm$^{-3}$ s at and behind the rim in the northwest and southwest
(regions 1, 2, 4, and 5) require a lower density closer to about
0.1-0.15 cm$^{-3}$; the higher ionization age of a few $10^9$
cm$^{-3}$ measured in the north and northeast (regions 6, 7, and NE)
would suggest higher ambient densities above 1 cm$^{-3}$.  For a
higher shock velocity, the expected change in ionization age would be
proportionally lower.

In our fits, higher ionization ages are also generally accompanied by
higher fitted column densities that are associated with the need to
absorb the He-like O lines predicted by the models.  These are
inconsistent, however, with the column density of $4.5 \times 10^{21}$
cm$^{-2}$ determined in the radio towards Tycho's SNR (Albinson et
al. 1986), and with the fitted column densities for the adjacent
regions on the rim.  This discrepancy most likely indicates spectral
complexity that is not well-represented by a single temperature and
ionization age for all the elements.  We were not able to resolve it
with any simple spectral models.

A third set of spectra taken directly behind the two sets already
discussed for rim segments 1, 2, 4, and 5 do finally show the
emergence of strong emission lines of Si and S associated with the
ejecta.  The ionization age limits for the forward shocked component
continue to be low at a few $10^8$ cm$^{-3}$ s for all except region
5.  The best two-component fit for region 5 shows a large increase in
the ionization age associated with
% factor of ten
the forward shock, but it also shows a large increase in the column
density, suggesting that the simple two-component model is not fully
adequate.

Results of the nonthermal fits are given in Table 5 for the prominent
major sections of the rim.  The quality of the fits is generally
comparable to, or perhaps slightly worse than the thermal model fits.
The ejecta are included in the model for the NE filament, following
the preceding discussion for the thermal fits.  The radio flux has
been fixed for the fits, but allowing the radio normalization to vary
between 15\% limits typically affects the fitted spectral index or
rolloff frequency by 20-30\%.  The spectrum of the northeast rim
is entirely consistent with the radio spectral index, giving a rolloff
frequency near 5$\times 10^{16}$ Hz that is slightly lower than that
estimated by Reynolds \& Keohane (1999) for the spatially integrated
spectrum of Tycho's SNR.  For the southwest, the fitted radio spectral
index is acceptable, but an improved fit can be obtained with a
slightly higher value that is closer to the overall average, and
entirely consistent with the uncertainties in the radio measurement.
Again, the rolloff frequency is then near 5$\times 10^{16}$ Hz.  The
situation in the northwest has been discussed above: the value of
the radio spectral index favored by the X-ray spectral fits is
marginally incompatible with the radio measurement, and the fitted
rolloff frequency is possibly slightly higher than in the other
sections of the rim that we examined.

The fitted values of the rolloff frequency $\nu$ in Table 5 can be
used to deduce the maximum electron energy $E$, given the strength of
the magnetic field $B$, using Eq. (2) of Reynolds \& Keohane (1999):
$$\nu (Hz) = 0.5 \times 10^{16} Hz \left( \frac{B}{10\, \mu G}\right)
\left( \frac{E}{10\ TeV}\right)^2.$$ Reynolds \& Ellison (1992)
estimate the magnetic field strength in Tycho's SNR to be
$10^{-4}-10^{-2}$ G.  Taking the rolloff frequencies from Table 5 and
the lower end of the values for the magnetic field, the maximum
electron energies are 8-14 TeV; using the upper end of the magnetic
field values, the maximum electron energies are 10 times lower, at
$\sim$ 1 TeV.

The continuum may actually be a mixture of thermal and nonthermal
emission, but it is difficult to disentangle these contributions from
these spectra alone.  The spectra may be fitted with such a hybrid
continuum model, but the fits are not substantially improved.  In the
example given in Table 3, the best fit is very strongly dominated by
the thermal component and the determination of the temperature is
therefore not adversely affected.  In other cases, there are two
statistically comparable local minima in the parameter space---one
where the nonthermal component dominates the flux, and the other where
the thermal component dominates the flux.  When the thermal component
dominates the flux, the temperature inferred for the thermal component
is not very different from the case with no nonthermal contamination.
When the nonthermal component dominates the flux, it is not surprising
that the temperature of the thermal component is more poorly
constrained, with errors that can be as much as two times larger.
Good fits may indeed sometimes be obtained with a hybrid model that is
dominated by the nonthermal component, but in these cases, the thermal
component takes on a low temperature and high ionization age in order
to fit the weak line emission from ejecta.  From our exploration of
the spectra, we conclude that it is not likely for the temperature of
the thermal component to be substantially higher than deduced using
the purely thermal models.

\subsection{Results for Hard Continuum Knots}

We also examined the spectra of two prominent knotty features present
to the east and the west in the hard X-ray continuum image.  The
spectra of both knots clearly show emission lines, and the western
knot (CKW) is strongly contaminated by the presence of Si and S
ejecta.  To fit the spectra, we used two-component NEI models to
represent solar abundance forward shocked gas and reverse shocked Si
and S ejecta, as discussed in the previous section and shown in Table
6 and Figure 7.  The fitted ionization age for the forward shock
component in CKW is significantly higher than for the rims, at a few
$10^9$ cm$^{-3}$ s, while the eastern knot (CKE) appears to favor a
low ionization age.

The eastern feature CKE is near, but not at, the eastern boundary of
the remnant, and the western feature CKW is well inside the western
boundary.  It is possible that these are similar to the rim features
that are seen along the west and south, except that by projection they
appear in the interior of the remnant.  Their higher surface
brightness in the hard continuum image is qualitatively consistent
with their higher fitted temperatures ($\sim$3 keV, compared to 2
keV), but their true nature remains elusive.  The CKW feature has a
higher fitted ionization age than the rest of the rim, and while this
can generally be explained by a higher density, the ambient density is
known to be higher in the east than it is in the west.  One would
therefore expect the opposite case of a higher ionization age for CKE
than for CKW.  If nonthermal emission were dominant, features
corresponding to the knots might be expected in the radio images, but
none are seen.

\section{Discussion}

\subsection{Electron Temperatures}

We combined the X-ray spectral results with the radio and X-ray
expansion velocities to estimate the ratio of electron to ion
temperatures behind the forward shock in Tycho's SNR.  The temperature
attained by particles as they pass behind the shock depends on their
mass according to the shock jump conditions as $kT \sim \frac{3}{16} m
v_s^2$, where $m$ is the particle mass and $v_s$ the shock velocity.
The different particle species will exchange energy through Coulomb
collisions and eventually attain a single equilibrium temperature, but
it has been proposed that this equilibration may be effected much
faster by collective plasma interactions (Cargill \& Papadopoulos
1988).  The efficiency of these collisionless heating processes
appears to decrease in inverse proportion to the Mach number of the
forward shock, however, so that they may be less important for the
fast shocks associated with young SNRs (Laming et
al. 1996, Ghavamian et al. 2001).  The electrons behind the slower
shocks in the Cygnus Loop, for example, have virtually equilibrated
with the ions, in contrast to those behind the faster shocks of
younger remnants such as SN 1006.

We plot in Figure 8 the equilibrium temperatures for Tycho's SNR
corresponding to the radio expansion parameters determined as a
function of azimuthal angle (defined counterclockwise from north) by
Reynoso et al. (1997).  We use their Table 2 and Figure 5, and a
distance of 2.3 kpc, to compute the velocities, and take the mean
molecular weight per particle to be $\mu m_p$, with $\mu=0.6$ for
solar abundance gas and $m_p$ being the proton mass. The rather large
fluctuations for angles greater than 180$^\circ$ are attributed by
Reynoso et al. (1997) to uncertainties in determining the remnant
radius.  Also shown in Figure 8 are the equilibrium temperatures
corresponding to the average radio expansion velocity, and the average
X-ray expansion velocity given by Hughes (2000).  It is not yet clear
which of the radio or X-ray expansion values should be used, though
the radio measurement is made with higher angular resolution data.
Also, if the ejecta presently located near the forward shock have high
velocities, they might bias the X-ray expansion measurement, as
suggested by Wang \& Chevalier (2001).  This question will hopefully
be better resolved with a new study of the X-ray proper motion using
higher spatial resolution data.

The data points in the Figure give the temperatures determined along
the rim from our fits, assuming that the X-ray emission is thermal
(Table 4); the appropriate azimuthal angle is determined with
reference to the center adopted by Reynoso et al. (1997).  The
electron temperatures we measure are clearly much lower than the mean
temperatures expected for shock velocities determined from the radio
expansion, with electron to mean temperature ratios of 0.10$-$0.20.
If we scale the mean temperature by the ratio of the average X-ray
velocity to the average radio velocity, it increases by a factor of
two, and the electron to mean temperature ratios decrease accordingly.
Although electron-ion temperature equilibration is not attained, the
electron temperatures are higher than would be expected from Coulomb
heating alone.  Ionization ages as low as $10^8$ cm$^{-3}$ s allow for
negligible electron heating mediated by Coulomb collisions.  The
measured electron temperature is thus essentially the electron
temperature attained immediately after passage through the shock
front, and represents extra heating in addition to Coulomb heating.
%with expected electron temperatures with Coulomb heating only are only
%about 0.1-0.2 keV.  

A low degree of electron-ion equilibration is consistent with results
from the optical spectra for the brightest knot in Tycho's SNR (knot
$g$), where the electron to proton temperature ratio is $\lesssim 0.1$
(Ghavamian et al. 2001).  The degree of electron-ion equilibration has
also been determined to be low in SN 1006, with the electron-to-proton
temperature ratio $\lesssim$ 0.05 using UV spectra (Laming et
al. 1996). In SN 1987A, the temperatures measured from Chandra X-ray
spectra are also lower than would be expected for equilibrium based on
the observed radio and X-ray expansion of the remnant (Michael et
al. 2002).  The implied ratio of electron and mean particle
temperatures is about 0.1, giving an electron to proton temperature
ratio of about 0.07 (with $\mu$ = 7 for the N-enriched circumstellar
material surrounding SN 1987A).  The X-ray line profiles (observed
with the high spectral resolution High Energy Transmission Gratings on
Chandra) are also consistent with thermal broadening due to high ion
temperatures behind the shock.  All these results are summarized in
Figure 9. For comparison, the model of Cargill \& Papadopoulos (1988)
for rapid electron heating can accommodate electron-to-ion temperature
ratios of about 0.20.

In contrast to these cases, the measured electron temperature behind
the forward shock in the Small Magellanic Cloud remnant E0102-72 is
lower than would be expected for the X-ray determined shock velocity,
even with Coulomb heating alone.  In this case, the low temperature is
interpreted as evidence for highly efficient, nonlinear acceleration
of particles behind the shock front (Hughes, Rakowski, \& Decourchelle
2000).  There is independent evidence that nonlinear acceleration of
electrons may be occuring in Tycho's SNR, in that the observed
curvature of the radio electron synchrotron spectrum is predicted by
such models (Reynolds \& Ellison 1992).  The brightness of the radio
emission at the rim also requires the fresh acceleration of particles
at the shock (Dickel et al. 1991), and this process may extend to
sufficiently high energies to affect the X-ray emission.  The
temperatures expected behind shocks where efficient particle
acceleration takes place are lower, making the ratio of measured
electron temperature to mean temperature ratios higher than otherwise.

For the CKE, the observed temperature is actually consistent with
temperature equilibration behind a shock at the radio velocity.  The
radio velocity of the rim may not be applicable there, however, as the
knot is interior to the rim and is probably only projected there.
This knot is also very near the position of the optical knot {\it g},
for which Ghavamian et al. (2001) determine that the electron
temperature should be no higher than 0.8 keV, at 0.1 times the proton
temperature.  However, there is a 10$''$ nominal offset between the
X-ray and optical (Kamper \& van den Bergh 1978) position, whereas the
nominal X-ray coordinate uncertainty should be well under 3$''$
(Chandra Proposer's Observatory Guide); the optical measurements are
probably not applicable to the continuum knot either.  One of the
radio velocity points applicable to rim segment 4 is also consistent
with the measured X-ray temperature, but this point represents a sharp
excursion well below the average.

The foregoing discussion is based on the assumption that the emission
is thermal, but it has been seen that nonthermal and thermal models
are about equally successful in individually describing the
featureless rim spectra.  In section 3.2, we argued that the spectra
indicate that the temperatures should not be much higher than
determined from the thermal fits alone, even in the presence of
nonthermal emission.  In the next section, we present evidence that an
additional nonthermal component is indeed likely to be present.

\subsection{Nonthermal and Thermal Emission Components}

Hard X-ray emission has been unmistakably detected from Tycho's SNR,
at energies up to 25 keV with HEAO-1 (Pravdo et \& Smith 1979), 20 keV
with Ginga (Fink et al. 1994); and 30 keV with RXTE (Petre et
al. 1999).  Several models have been proposed to explain the hard
X-ray continuum in Tycho's SNR as synchrotron radiation from electrons
accelerated at the shock (Heavens 1984, Ammosov et al. 1994).  Given
that the temperatures (in thermal models) at the forward shock in
Tycho's SNR are only about 2 keV, the X-ray emission at higher
energies may well have a nonthermal origin, as has been suggested in
the literature.

We estimate an upper limit to the nonthermal X-ray luminosity from
Tycho's SNR in the Chandra energy band by attributing {\it all} the
flux between 4-6 keV to a nonthermal component.  This is clearly an
overestimate because there is a hot thermal component associated with
the Fe K line emission (Hwang et al. 1998).  We first obtained the
total source counts between 4-6 keV by estimating and subtracting the
background counts from the continuum image and multiplying by a rough
geometrical correction factor of 1.15 for the portion of the remnant
that was not imaged.  We then took the 0.5-10 keV source luminosity of
the nonthermal model for the northwest rim and scaled it up by the
ratio of 4-6 keV counts in the total image to those in the rim
segment, to estimate the nonthermal luminosity for the entire remnant.
This maximum luminosity is about 2 $\times 10^{35}$ ergs/s, but could
be reduced by a factor of a few if we use only the bright X-ray
continuum regions in the rim and the knots in our estimate.  We have
not accounted for the variation in radio spectral index for different
parts of the remnant, but if we use the spectrum fitted to the
southwest rim, which has a much flatter radio slope, the estimated
luminosity is essentially unchanged.

One can also estimate this nonthermal luminosity for Tycho's SNR by
extrapolating the flat component in the Ginga or RXTE X-ray spectra
down to energies in the Chandra band.  Extrapolating the power-law
model given by Fink et al. (1994) for the Ginga data, we estimate a
0.5-10 keV nonthermal luminosity of about 1.5$\times 10^{35}$
erg/s. Aharonian et al. (2001) fitted the RXTE spectrum with a model
that includes an exponential cutoff, although they inferred a steeper
radio slope than is actually observed overall, and obtained a higher
turnover energy of 1.6 keV ($\nu = 3.8 \times 10^{17}$ Hz).  The
0.5-10 keV luminosity implied by their model is $1 \times 10^{35}$
ergs/s.  Both estimates are a substantial fraction of the maximum
possible luminosity from our estimate above, indicating that perhaps
half of the hard 4-6 keV X-ray emission in Tycho's SNR is nonthermal.
The nonthermal luminosity estimated from these hard X-rays is slightly
higher than that of SN 1006 in the same energy range (Dyer et
al. 2001).

The nonthermal X-ray emission should also be accompanied by
$\gamma$-ray emission, since the energetic electrons that emit the
X-ray synchrotron radiation will also upscatter cosmic background
photons.  The $\gamma$-ray upper limits for Tycho's SNR have become
tighter in recent years (Aharonian et al. 2001), at levels that are a
few to several times lower than the detections for SN 1006 (Tanimori
et al. 1998).  This may be understandable from a consideration of the
densities and magnetic fields in these remnants.  The extrapolation of
the radio spectrum to the X-ray range gives maximum electron energies
comparable to those determined for SN 1006 (Reynolds \& Keohane 1999),
if equal magnetic field strengths of 10 $\mu$G are assumed.  A
magnetic field of $10^{-4}$ to $10^{-2}$ G in Tycho's SNR from
Reynolds \& Ellison (1992) is higher than the 9 $\mu$G in SN 1006
(Dyer et al. 2001).  A higher magnetic field qualitatively accounts
not only for the more severe synchrotron energy losses of the
electrons in Tycho's SNR, but also its higher radio surface
brightness.  The maximum electron energies will be lower in proportion
to $B^{-1/2}$; on the basis of our spectral fits, the maximum electron
energies would appear to be between 1-15 TeV, depending on the exact
value of the magnetic field.  The much lower density environment of SN
1006 (e.g., see Kirshner et al. 1987) would also allow particle
acceleration to proceed to higher energies in the first place (Baring
et al. 1999).

\subsection{Mixing of Ejecta}

The presence of ejecta at the forward shock requires a significant
amount of mixing either during or after the explosion.  Various high
resolution hydrodynamical studies (Dwarkadas 2000, Wang \& Chevalier
2001) show that the ejecta cannot penetrate beyond about half the
interaction region through Rayleigh-Taylor instabilities alone, though
early Richtmyer-Meshkov instabilities might enhance the effectiveness
of the Rayleigh-Taylor instabilities (Kane et al. 1999).  Clumps in
the ejecta (Hamilton 1985) or in the interstellar medium (Jun, Jones
\& Norman 1996) would, however, allow the ejecta to penetrate to large
radii.  Ejecta might also penetrate the forward shock in the presence
of enhanced turbulence generated as the reverse shock propagates into
low density bubbles of Fe, formed as the heat release from the
radioactive decay of clumps of Ni ejecta allows them to expand
(Blondin et al. 2001).  Yet another possibility is that the ejecta can
affect the forward shock when the interaction region for the shock is
very thin, as could be the case with nonlinear particle acceleration
behind the forward shock---a situation that might reasonably be
expected to occur in young remnants (Blondin \& Ellison 2001),
including Tycho's SNR in particular (Reynolds \& Ellison 1992).  The
action of the Rayleigh-Taylor instabilities out to the forward shock
is further supported by the radial orientation of the magnetic field
(Dickel et al. 1991), which is observed all the way to the rim of the
remnant.

One might expect that if the ejecta do penetrate the forward shock,
the boundary of the outer shock would be distorted, whereas the
western boundary of Tycho's SNR is smooth and circular.  This might
still be possible, however, if the mixing occured early enough for the
protruberances to have subsided.  In their simulations, Wang \&
Chevalier (2001) found that it was actually difficult to deform the
forward shock unless the ejecta density contrast was very high.

The qualititative appearance of the Si ejecta in Tycho's SNR is quite
different from that of the core-collapse remnant Cas A, which has also
been beautifully imaged by Chandra (Hughes et al. 2000, Hwang et
al. 2000).  While the Si is clumpy in Tycho, it is compact and knotty
in Cas~A.  Despite the fact that Tycho is located at a smaller
distance, the angular scale of its Si features is larger than those
seen in Cas A.  Differences are also seen in the 4-6 keV continuum
images.  In Tycho's SNR, the ejecta shell has virtually disappeared in
this image, but in Cas A, the ejecta are still clearly visible.  The
emission associated with the ejecta is thus seen to be relatively more
important at these energies in Cas A than in Tycho's SNR.  Indeed, the
hard X-ray emission in Cas A has been proposed to be primarily
bremsstrahlung emission from electrons that have been accelerated in
the ejecta (Laming 2001ab, Bleeker et al. 2001).

The appearance of the forward shock in these two remnants is also
strikingly different.  Tycho's outer rim is distorted because of its
interaction in the east, but the outline of the rim is otherwise
generally smooth and continuous.  By contrast, the outer rim of Cas A
is broken up on small scales and shows tightly curved fragments
(Gotthelf et al. 2000).  If some of this emission is nonthermal, this
may reflect differences seen in their radio emission, as Tycho has a
sharp radio rim suggesting efficient first order Fermi acceleration,
whereas Cas A has no distinct radio rim and is a better candidate for
second order Fermi acceleration mediated by turbulence (e.g., Dickel
et al. 1991).  SN 1006, another Type Ia remnant, shows the same smooth
outer rim seen in Tycho's SNR, and the same highly clumpy ejecta (in
this case O ejecta; K. Long, private communication), but this is still
much too small a sample for these differences to be more than
suggestive.

\section{Summary}

The Chandra observatory allows the imaging and spectral study of the
forward shock in Tycho's SNR.  The Si ejecta are highly clumpy and
have propagated to the forward shock in several locations.  The
spectra at the rim show slight, but varying amounts of line emission,
with continua that are well described by either thermal or nonthermal
models.  Spectra behind the shock show stronger line emission
consistent with this gas having been shocked at earlier times.  The
limits on the change in ionization age behind the rim give some
indication of variations in density around the remnant. Taking the
electron temperature from thermal models, combined with the results of
radio and X-ray expansion studies, the electron to mean temperature
ratios behind the shock are $\lesssim 0.1-0.2$, indicating a modest
amount of electron heating, but not full temperature equilibration.  A
consideration of the hard X-ray emission suggests, however, that an
additional nonthermal component must be present.  Even in the presence
of such a nonthermal component, the electron temperatures do not seem
likely to significantly exceed the values we measure assuming purely
thermal models.  Ideally, it is desirable to have energy coverage that
continues above 10 keV to disentangle the nonthermal and thermal
contributions to the spectrum, but such observations will not be
possible in the near future.

For future work, a useful first step would be imaging the distribution
of the high energy X-ray flux above 8 keV, as would be possible with
XMM-Newton.  To the extent that the hard X-rays are distributed like
the 4-6 keV X-rays at the rim of the remnant (as seems likely), this
would be consistent with their origin from electrons accelerated to
high energies.  It would also be beneficial to model as wide a
bandwidth spectrum as possible, and to carry out a detailed study of
the radio and X-ray correlations, as well as to consider the effects
of nonlinear particle acceleration.

\acknowledgments 

We are grateful to John Dickel for generously providing his
flux-normalized 22 cm radio image, Ramesh Narayan for an inspiring
conversation, and Andrew Szymkowiak for several helpful scientific
discussions.  We also thank our anonymous referee for a very careful
review of the manuscript.

\begin{figure}
%\centerline{\includegraphics[scale=0.40]{f1a_bw.ps}\hspace{0.2in}\includegraphics[scale=0.40]{f1b_bw.ps}}
%\centerline{\includegraphics[scale=0.40]{f1c_bw.ps}\hspace{0.2in}\includegraphics[scale=0.40]{f1d_bw.ps}}
\centerline{\includegraphics[scale=0.8]{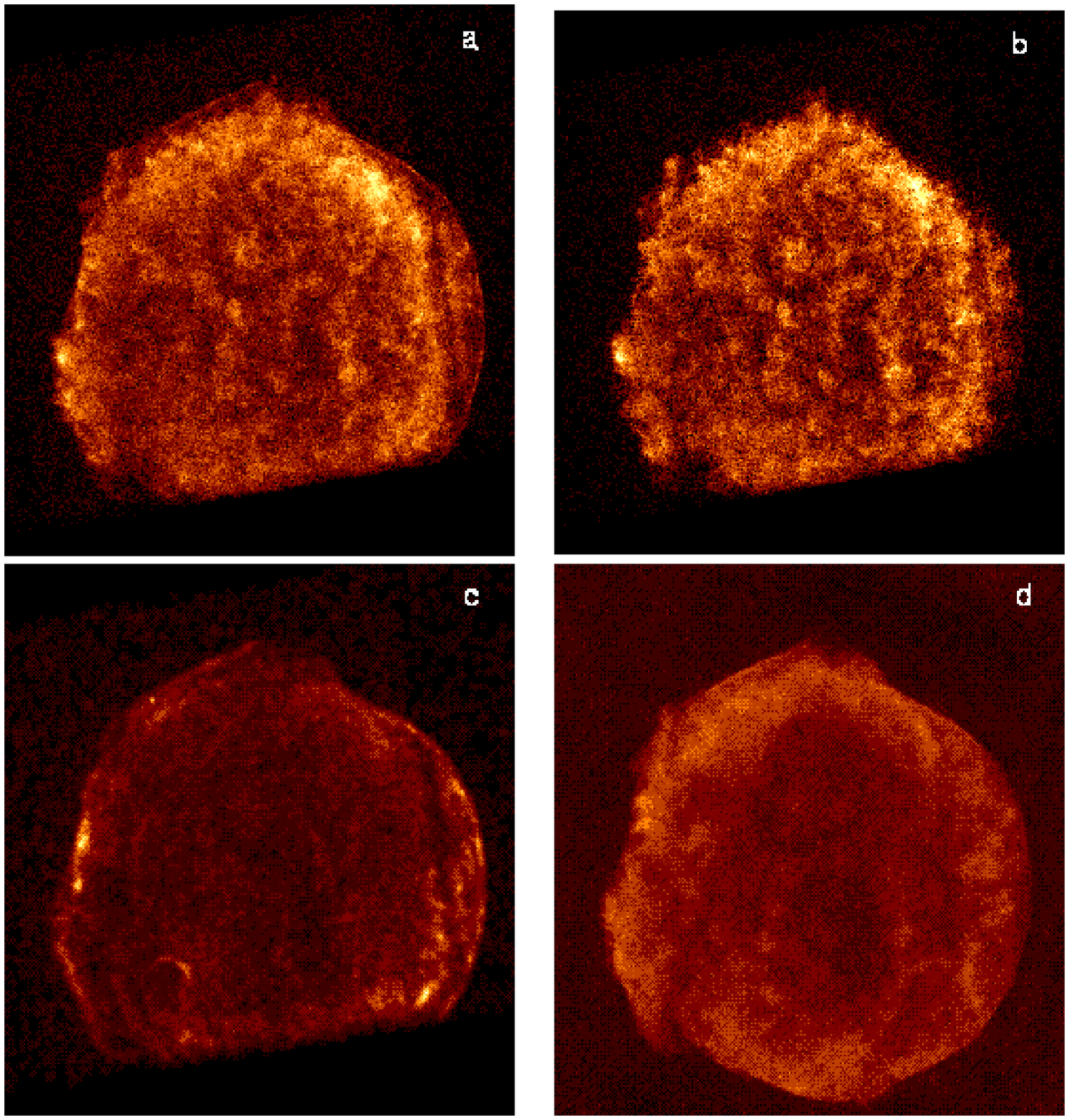}}

\caption{(a) Unsmoothed broadband Chandra ACIS image of Tycho's SNR.
Most of the remnant is imaged on S3, but the western portion falls on
S2 (the chip gap is visible), and the southernmost portion is not
imaged.  (b) Unsmoothed image of the Si emission lines near 1.86 keV
with the underlying continuum subtracted.  (c) Smoothed image of the
hard continuum region at energies between 4-6 keV. (d) Radio image at
22 cm (1.36 GHz, epoch 1983; courtesy of John Dickel), convolved with
a beam of width $1.45'' \times 1.38''$ and scaled to match the X-ray
images in size.  All images have a square-root intensity scaling,
except the radio image, which has a linear scaling.  North is up and
east to the left.  The remnant is roughly 8$'$ across.}
\end{figure}

\begin{figure}
%\centerline{\includegraphics[scale=0.5]{f2a.ps}\hspace{-0.06in}
%\includegraphics[scale=0.50]{f2b.ps}\hspace{-0.06in}
%\includegraphics[scale=0.50]{f2c.ps}}
\centerline{\includegraphics[scale=0.8]{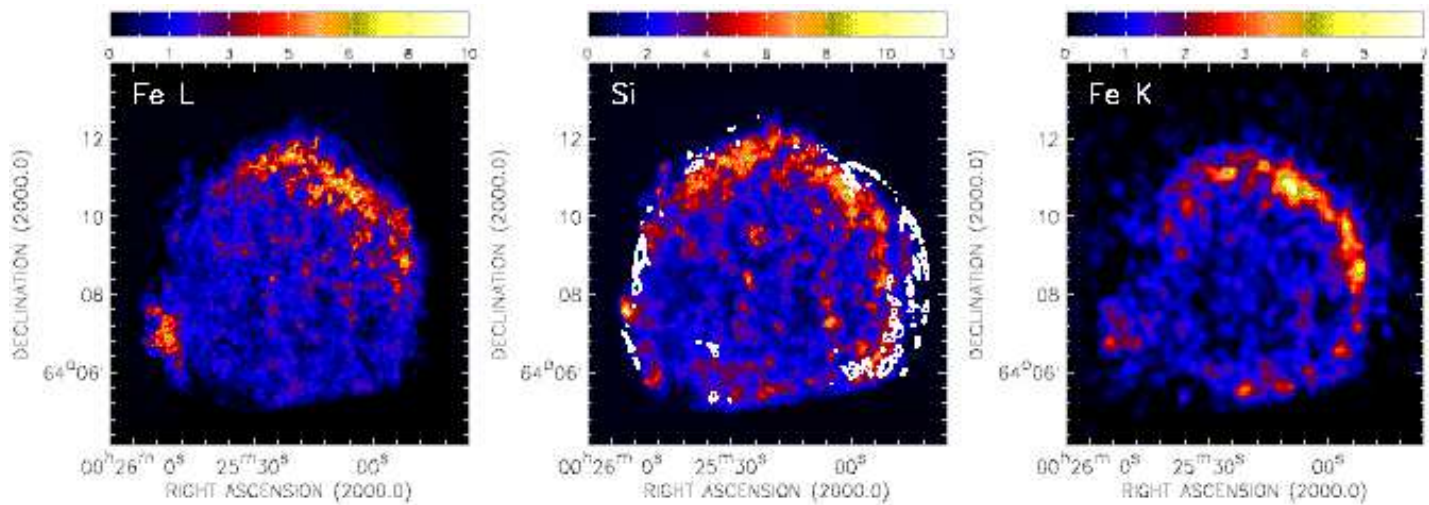}}
\caption{Equivalent width (continuum-subtracted, line-to-continuum
ratio) image of (a, left) Fe L emission (n=2 transitions near 1 keV)
(b, middle) Si emission (transitions of He- and H-like ions near 1.86
keV and 2.006 keV), with low-level (10 and 20\% maximum) contours of
the smoothed continuum emission overlaid (c, right) Fe K emission (n=1
transitions near 6.5 keV).}
\end{figure}

\begin{figure}
\centerline{\includegraphics[scale=0.4,angle=90]{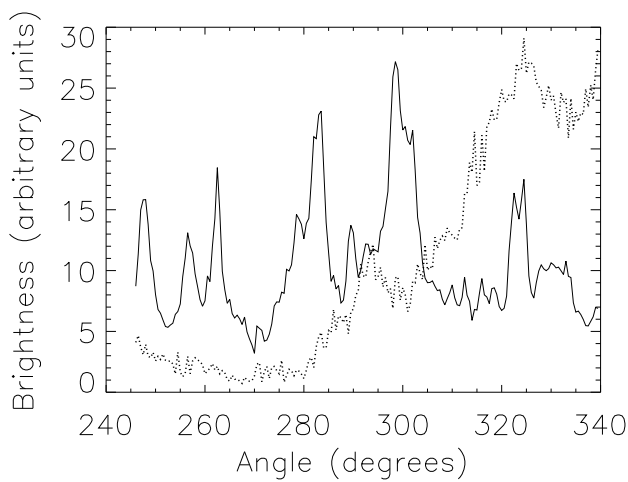}}
\caption{Comparison of X-ray continuum (solid) and radio (dotted)
brightness along a 9$''$ wide circular arc of radius 220$''$, plotted
against azimuthal angle measured counter-clockwise from north.  One
degree of azimuth corresponds to 3.8$''$ angular distance along the
arc.}
\end{figure}

\begin{figure}
\centerline{\includegraphics[scale=0.50,angle=90]{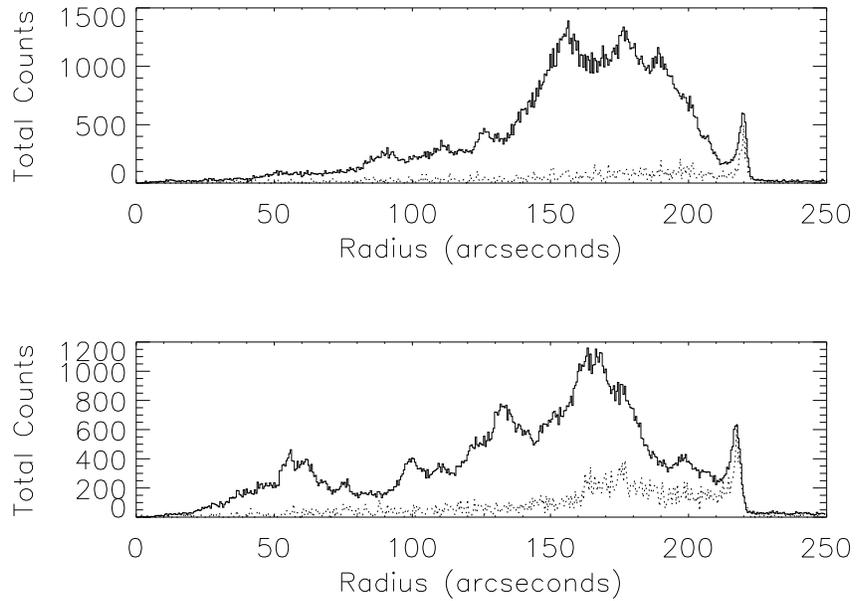}}
\caption{Radial profiles ending in portions of the northwestern (top;
towards region 5) and southwestern (bottom; towards region 1) rims.
The solid lines show the profiles of the broadband image, and the dotted
lines show the profiles of the continuum image, multiplied by a factor
of 15. }
\end{figure}

\begin{figure}
%\centerline{\includegraphics[scale=0.4]{f5.ps}}
\centerline{\includegraphics[scale=0.4]{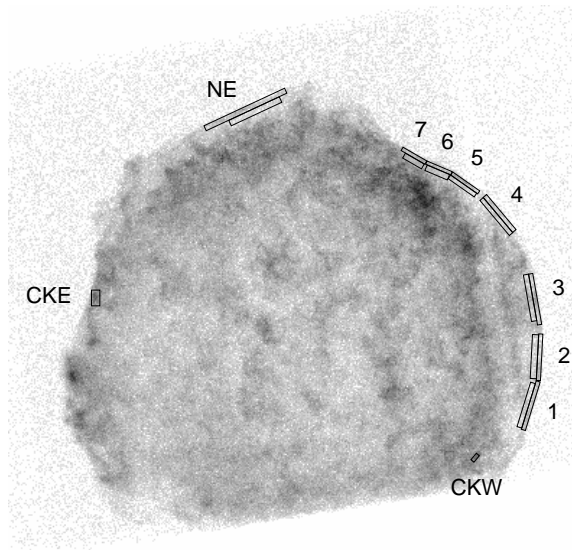}}
\caption{Chandra broadband image with the spectral extraction regions overlaid and labelled.}
\end{figure}

\begin{figure}
\centerline{\includegraphics[scale=0.3,angle=-90]{f6a.ps}\includegraphics[scale=0.3,angle=-90]{f6aa.ps}}
\centerline{\includegraphics[scale=0.3,angle=-90]{f6b.ps}\includegraphics[scale=0.3,angle=-90]{f6bb.ps}}
\centerline{\includegraphics[scale=0.3,angle=-90]{f6c.ps}\includegraphics[scale=0.3,angle=-90]{f6cc.ps}}
\centerline{\includegraphics[scale=0.3,angle=-90]{f6d.ps}\includegraphics[scale=0.3,angle=-90]{f6dd.ps}}
\caption{}
\end{figure} \setcounter{figure}{5}\begin{figure}
\centerline{\includegraphics[scale=0.3,angle=-90]{f6e.ps}\includegraphics[scale=0.3,angle=-90]{f6ee.ps}}
\centerline{\includegraphics[scale=0.3,angle=-90]{f6f.ps}\includegraphics[scale=0.3,angle=-90]{f6ff.ps}}
\centerline{\includegraphics[scale=0.3,angle=-90]{f6g.ps}\includegraphics[scale=0.3,angle=-90]{f6gg.ps}}
\centerline{\includegraphics[scale=0.3,angle=-90]{f6h.ps}\includegraphics[scale=0.3,angle=-90]{f6hh.ps}}
\caption{}
\end{figure} \setcounter{figure}{5}\begin{figure}
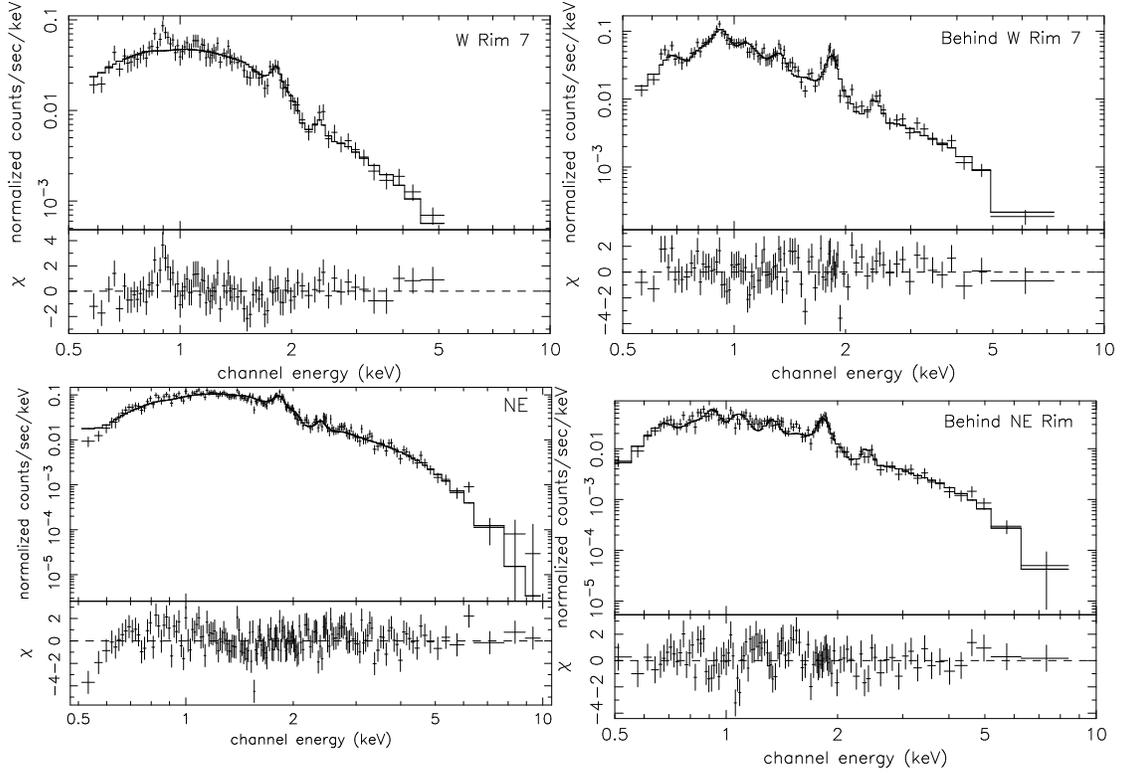

\centerline{\includegraphics[scale=0.3,angle=-90]{f6i.ps}\includegraphics[scale=0.3,angle=-90]{f6ii.ps}}
\centerline{\includegraphics[scale=0.3,angle=-90]{f6j.ps}\includegraphics[scale=0.3,angle=-90]{f6jj.ps}}
%\centerline{\includegraphics[scale=0.3,angle=-90]{f6k.ps}\includegraphics[scale=0.3,angle=-90]{f6kk.ps}}
\caption{Spectra at and behind the rim shown with fitted thermal NEI models described in the text, starting from the SW (see Fig. 3) and progressing counterclockwise to the NE.}
\end{figure}

\begin{figure}
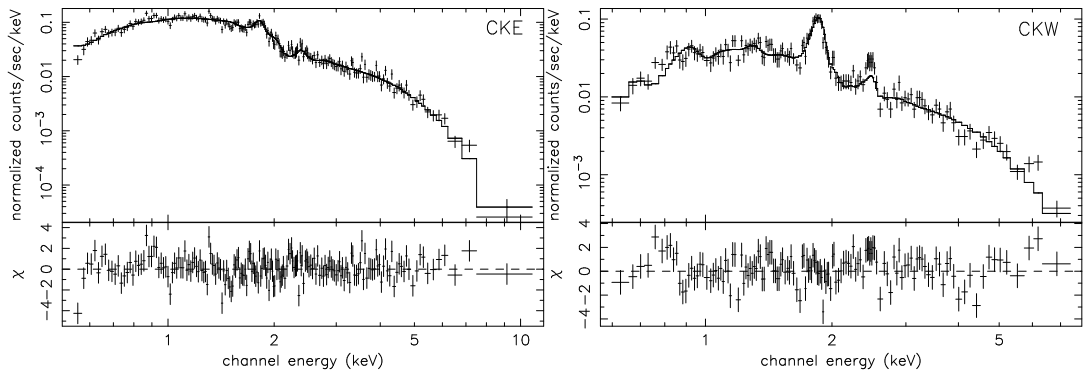

\centerline{\includegraphics[scale=0.3,angle=-90]{f7a.ps}\includegraphics[scale=0.3,angle=-90]{f7b.ps}}
\caption{Spectra of the eastern and western hard continuum knots, with
two-component NEI models representing forward shocked gas and reverse shocked
ejecta.}
\end{figure}

\begin{figure}
\plotone{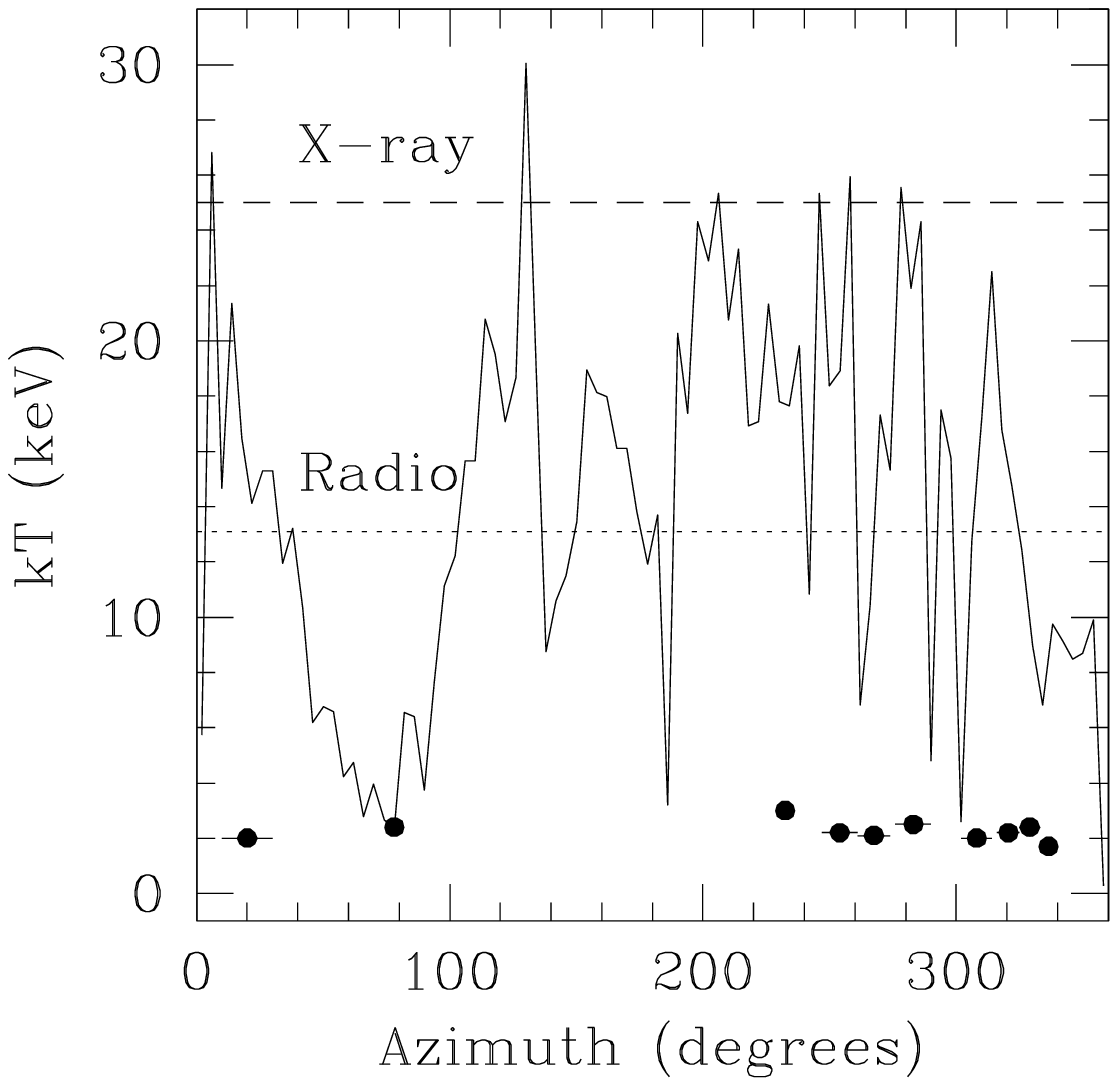}
\caption{Equilibrium mean temperatures behind the shock front from
radio shock velocities computed from Reynoso et al. (1997), plotted
against azimuthal angle increasing counterclockwise from north.  The
scatter in the south and southwest (angles greater than 200) is
attributed to uncertainties in the radio expansion from uncertainties
in obtaining the radius.  The dotted line shows the temperature
corresponding to the mean radio expansion, and the dashed line shows
the temperature corresponding to the mean X-ray expansion from Hughes
(2000).  The measured electron temperatures at the rim from Table 4 and
the two hard knots from Table 5 are plotted as the data points
starting with the NE, CKE, CKW, and on through the rim regions 1
through 7.  The temperature error bars are smaller than the data
points, and the angular extent of the spectral extraction region is
indicated by the horizontal bars.}
\end{figure}

\begin{figure}
%\plotone{f9.ps}
\plotone{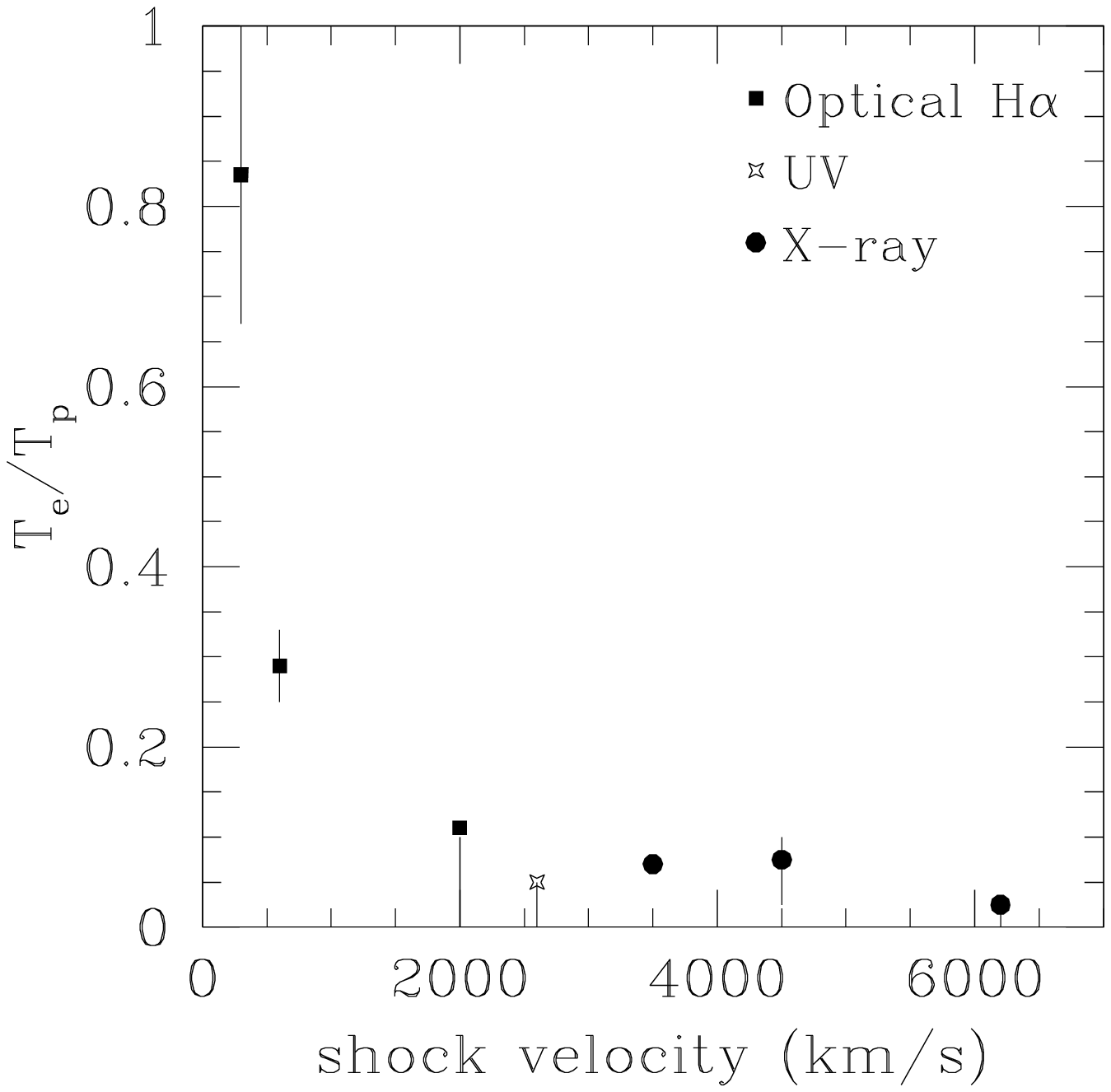}
\caption{Ratios of electron and proton temperatures inferred for
supernova remnant shocks at various wavelengths.  Given in order of
increasing shock velocity, optical results for nonradiative shocks in
the Cygnus Loop, RCW 86, and Tycho's SNR are shown as the squares and
are taken from Ghavamian et al. (2001); the UV result for SN 1006 are
shown as a star and are from Laming et al. (1996); the X-ray results
are shown as circles for SN 1987A from Michael et al. (2002) using
mean mass per particle of 0.7; for Tycho's SNR from this work, and
E0102-72 from Hughes et al. (2000) using a mean mass per particle of
0.6.}
\end{figure}

\begin{deluxetable}{lll}
%\tabletypesize{\scriptsize}
\tablecaption{Energy Band Images }
\tablewidth{0pt}
\tablehead{
\colhead{Image} & \colhead{Pulse Height$^*$} & \colhead{Raw Counts} }
\startdata
Total     & 1-1024  & 5.5$\times 10^6$ \\
Fe L      & 46-65   & 1.3$\times 10^6$ \\
Si        & 112-135 & 1.8$\times 10^6$ \\
Continuum & 276-424 & 50,000 \\
Fe K      & 428-451 & 7,000 \\
\enddata
\tablenotetext{*}{Nominal energies are obtained from the pulse
heights, which are corrected for the spatial variation in the gain,
as energy in eV =  pulse height $\times$ 14.6.}
\end{deluxetable}

\begin{deluxetable}{llll}
\tabletypesize{\scriptsize}
\tablecaption{Regions}
\tablewidth{0pt}
\tablehead{
\colhead{Region}  & \colhead{Cnts} & \colhead{Length} & \colhead{Width} \\
\colhead{}  & \colhead{} & \colhead{($0.5''$ pixels)} & \colhead{($0.5''$ pixels)} }
\startdata
Rim 1 (SW) & 4281 & 110 & 9 \\
Behind 1 (SW)& 2587 & 100 & 11\\
Rim 2 (SW) & 3379 & 101 & 9 \\
Behind 2 (SW)& 3045 & 95 & 13 \\
Rim 3      & 5020 & 112 & 9 \\
Behind 3   & 4600 & 102 & 11\\
Rim 4 (NW) & 2829 & 99 & 10 \\
Behind 4 (NW)& 2732 & 99 & 10\\
Rim 5 (NW) & 3568 & 71 & 8 \\
Behind 5 (NW)& 1834 & 69 & 10\\
Rim 6      & 2415 & 59 & 8 \\
Behind 6   & 3269 & 53 & 13 \\
Rim 7      & 3249 & 59 & 8 \\
Behind 7   & 3760 & 47 & 12 \\
Rim NE     & 8171 & 192 & 12 \\
Behind NE  & 3228 & 120 & 13 \\
\enddata
\end{deluxetable}

\begin{deluxetable}{llllll}
\tabletypesize{\scriptsize}
\tablecaption{Northwest Rim (4+5): Spectral Fits}
\tablewidth{0pt}
\tablehead{
\colhead{Model}  & \colhead{$\chi^2, \chi^2/dof$} & \colhead{$N_H$} & \colhead{$kT$ or $\alpha^a$} & \colhead{$n_et$ or $\nu^b$} & \colhead{Normalization}  \cr
\colhead{}  & \colhead{} & \colhead{($10^{22}$ cm$^{-2}$)} & \colhead{(keV or --)} & \colhead{(cm$^{-3}$ s or Hz)} & \colhead{}  
}
\startdata
bremsstrahlung & 179.9, 1.28 & 0.53 (0.50-0.56) & 2.1 ( 2.0-2.2) & -- & 6.2e-4 (5.8-6.7e-4) \\
parallel shock & 179.6, 1.28 & 0.53 (0.50-0.56) & 2.1 (1.9-2.2) & 1.1e8 ($<$3.5e8) & 1.9e-3 (1.8-2.1e-3) \\
NEI            & 179.6, 1.28 & 0.53 (0.50-0.56) & 2.1 (1.9-2.2) & 1.0e8 ($<$2.3e8) & .. \\
power-law      & 217.9, 1.55 & 0.75 (0.71-0.79) & 2.79 (2.71-2.87) & -- & 7.8e-4 (7.1-8.4e-4) \\
srcut          & 199.4, 1.41 & 0.66 (0.63-0.69) & 0.52 (0.51-0.53) & 7.0e16 (5.4-8.4e16) & [0.086]$^c$ \\
               & 201.5, 1.43 & 0.68 (0.65-0.70) & [0.72] & 1.1e17 (0.90-1.4e17) & 2.8 (2.5-3.6) \\
srcut+NEI      & 179.6, 1.29 & 0.53 (0.52-0.62) & $\alpha$=[0.52] & $\nu$=1.4e15 (1.0e10-2.5e16) & [0.086] \\
               &             &      & $kT$=2.1 (1.9-2.2) & $nt$=1.0e8 ($<$1.8e8) & 1.9e-3 (1.5-2.1e-3; $>$75\% of flux)\\

\enddata \tablenotetext{a}{Temperature $kT$ is given for
bremsstrahlung, parallel shock, and NEI models, spectral index
$\alpha$ in X-rays for power-law, and in radio for $srcut$ models.
Please see the text for further discussion.}
\tablenotetext{b}{Ionization age $n_et$ is given for planar shock and
NEI models, rolloff frequency $\nu$ for nonthermal $srcut$ models.}
\tablenotetext{c}{In this and other tables, quantities in square
brackets are held fixed at the value given.}
\end{deluxetable}

\begin{deluxetable}{llllllll}
\tabletypesize{\scriptsize}
\tablecaption{Rim: Spectral Fits with Thermal $NEI$ Models}
\tablewidth{0pt}
\tablehead{
\colhead{Region}  & \colhead{$\chi^2, \chi^2/dof$} & \colhead{$N_H$} & \colhead{FS: $kT$} & \colhead{$n_et$} & \colhead{RS$^\dagger$: $kT$} & \colhead{$n_et$}  & \colhead{Fe}  \cr
 \colhead{}  & \colhead{} & \colhead{($10^{22}$ cm$^{-2}$)} & \colhead{(keV)} & \colhead{(cm$^{-3}$ s)} & \colhead{(keV)} & \colhead{(cm$^{-3}$ s)} & \colhead{(rel $\odot$)}  
}
\startdata
Rim 1        & 135.8, 1.18 & 0.60 & 2.2       & 1.0e8      & --   & --     & --  \\
             &      & (0.56-0.64) & (2.0-2.3) & ($<$2.2e8) &      &        &     \\
Behind 1     & 68.5, 0.94  & 0.46 & 1.9       & 1.0e8      & [0.9]& 1.2e11 & [0]  \\
             &      & (0.42-0.52) & (1.7-2.1) & ($<$2.5e8) &      & (3.4e10-9.0e11) &     \\
Rim 2        & 140.6, 1.42$^\ddagger$ & 0.63 & 2.1       & 3.2e8      & --  & -- & --  \\
             &      & (0.57-0.68) & (2.0-2.3) & ($<$4.3e8) &      &       &     \\
Behind 2     & 95.7, 1.10  & 0.51 & 2.0       & 2.5e8      & --   & --     & --  \\
             &      & (0.45-0.57) & (1.8-2.2) & ($<$3.5e8) &      &        &     \\
SW Rim (1-2) & 206.7, 1.18 & 0.61 & 2.1       & 1.0e8      & --   & --     & --  \\
             &      & (0.58-0.65) & (2.0-2.2) & ($<$2.2e8) &      &        &     \\
Behind SW Rim& 209.2, 1.44 & 0.51 & 1.9       & 1.8e8      & --   & --     & --  \\
             &      & (0.47-0.55) & (1.8-2.0) & ($<$2.6e8) &      &        &     \\
Rim 3        & 150.0, 1.19 & 0.48 & 2.6       & 1.0e8      &[0.9] & 1.4e11 & [0] \\
             &      & (0.44-0.51) & (2.4-2.8) & ($<$2.0e8) &      &($<$ 6.9e11) &     \\    
Behind 3$^*$ & 132.8, 1.24 & 0.52 & 2.0       & 1.0e8      &[0.9] & 8.1e10 & 0.04 \\
             &      & (0.47-0.62) & (1.8-2.2) & ($<$2.1e8) &      & (5.6e10-1.1e11) & (0.03-0.05) \\
Rim 4        & 89.1, 1.13  & 0.57 & 2.0       & 1.0e8      & --   & --     & --              \\
             &      & (0.52-0.63) & (1.8-2.1) & ($<$2.7e8) &      &        &                 \\
Behind 4     & 108.3, 1.39$^\ddagger$ & 0.46 & 2.0       & 1.8e8      & --   & --     & --              \\
             &      & (0.41-0.52) & (1.8-2.2) & ($<$3.1e8) &      &        &                 \\
Rim 5        & 97.0, 0.95  & 0.49 & 2.2       & 1.0e8      & --   & --     & --              \\
             &      & (0.45-0.53) & (2.0-2.4) & ($<$2.5e8) &      &        &                 \\
Behind 5     & 51.7, 1.08  & 0.29 & 1.6       & 1.0e8      & --   & --     & --              \\
             &      & (0.26-0.34) & (1.4-1.9) & ($<$1.7e8) &      &        &                 \\
NW Rim (4-5) & 179.4, 1.28 & 0.53 & 2.1       & 1.0e8      & --   & --     & --              \\
             &      & (0.50-0.56) & (1.9-2.2) & ($<$2.2e8) &      &        &                 \\
Behind NW Rim& 145.4, 1.25 & 0.39 & 1.9       & 1.0e8      & --   & --     & --              \\
             &      & (0.36-0.42) & (1.8-2.1) & ($<$3.0e8  &      &        &                 \\
Rim 6        & 84.9, 1.31 & 0.28 & 2.4 & 1.0e8 & [0.9] & 1.6e11 & [0] \\
             &      & (0.24-0.31) & (2.1-2.8) & ($<$2.1e8) &      & ($>$1.2e10)        &   \\
             & 69.2, 1.08 & 0.52 & 2.2 & 4.3e8 & [0.9] & 4.9e10 & 0.27 \\
             &      & (0.38-0.74) & (1.8-2.5) & (2.6-6.4e8) &      & (3.3-7.3e10)        &  (0.11-0.92) \\
Behind 6     & 104.7, 1.23 & 0.84 & 2.5       & 2.5e9      & [0.9]& 6.2e10 & 1.2             \\
             &      & (0.82-0.88) & (2.3-2.9) & (1.9-3.0e9)&      & (4.8-8.3e10) & (0.8-2.1) \\
%NW Rim (5-6) & 193.5, 1.30 & 0.43 & 2.2       & 1.0e8      & --   & --     & --              \\
%             &      & (0.41-0.46) & (2.0-2.3) & ($<$1.4e8) &      &        &                 \\
%Behind NW    & 172.5, 1.40 & 0.92 & 2.1       & 2.7e9      & [1.5]& 2.5e10 & 1.8             \\
%             &      & (0.88-0.95) & (1.9-2.3) & (2.3-3.2e9)&      &(2.1-3.2e10) & (1.1-2.7)  \\
Rim 7        & 96.8, 1.17 & 0.24 & 1.7 & 1.0e8 & [0.9] & 5.3e9 & [0] \\
             &      & (0.21-0.27) & (1.5-1.9) & ($<$1.4e8) &      & (3.2-7.7e9)       &                 \\
Behind 7     & 119.0, 1.31 & 0.88 & 2.0       & 3.3e9      &[0.9] & 8.9e10       & 0.8      \\
             &      & (0.85-0.91) & (1.8-2.2) & (2.9-3.7e9)&      & (6.5e10-1.3e11) & (0.6-1.1) \\
NE Rim       & 202.9, 1.23 & 0.44 & 2.0       & 1.0e8      & [0.9]& 6.3e9        & [0]       \\
             &      & (0.42-0.46) & (1.9-2.1) & ($<$1.2e8) &      & (4.5-8.8e9)  &           \\
Behind NE Rim & 116.9, 1.16 & 0.95 & 2.1       & 3.5e9      & [0.9]& 4.8e10       & [0]       \\
             &      & (0.92-0.97) & (1.9-2.3) & 3.2-3.9e9) &      & 1.6e10-4.9e11   &           \\
\enddata
\tablenotetext{\dagger}{Please see the text for discussion of the reverse shocked ejecta component.}
\tablenotetext{\ddagger}{Addition of an ejecta component improves the fit substantially, but is poorly constrained.}
\tablenotetext{*}{The S abundance was freely fitted to be 2.1 (1.7-2.5) relative to its solar value.}
\end{deluxetable}

\begin{deluxetable}{lllllll}
\tabletypesize{\scriptsize}
\tablecaption{Rim: Spectral Fits with Nonthermal $srcut$ Models}
\tablewidth{0pt}
\tablehead{
\colhead{Region}  & \colhead{$\chi^2, \chi^2/dof$} & \colhead{$N_H$} & \colhead{Fitted $\alpha$} & \colhead{Radio $\alpha^a$} & \colhead{Radio Flux$^b$}  & \colhead{$\nu^c$} \cr
\colhead{}  & \colhead{} & \colhead{($10^{22}$ cm$^{-2}$)} & \colhead{} & \colhead{} & \colhead{(Jy)}  & \colhead{(Hz)}
}
\startdata
NE$^d$  & 211.2, 1.27 & 0.58 (0.57-0.60) &[0.55] & 0.55$\pm$0.04 &  [0.21] &  5.2e16 (5.1-5.3 e16) \\
NW (4\&5) & 199.4, 1.41 & 0.66 (0.63-0.69) & 0.52 (0.51-0.53) & 0.72$\pm$0.14 & [0.086] & 7.0e16 (5.4-8.4e16) \\
%NW (5\&6) & 205.7, 1.37 & 0.54 (0.53-0.56) & 0.56 (0.55-0.57) & 0.72$\pm$0.14 & [0.10] & 9.6e16 (8.0-13e16) \\
SW (1\&2) & 259.5, 1.47 & 0.87 (0.85-0.90) &[0.40] & 0.40$\pm$0.15 & [0.03] & 3.3e16 (3.2-3.4e16) \\
 & 228.2, 1.30 & 0.76 (0.73-0.78) &0.440$\pm$0.005 &  0.40$\pm$0.15 & [0.03] & 5.8e16 (5.0-7.1e16) \\
\enddata
\tablenotetext{a}{Radio spectral index $\alpha$ as given by Katz-Stone et al. (2000).}
\tablenotetext{b}{Radio flux at 1 GHz (please see the text).}
\tablenotetext{c}{Rolloff frequency of the spectrum.}
\tablenotetext{d}{The model for this spectrum also includes an ejecta
component with kT fixed at 0.9 keV and $n_et$ = 7.6e9 (5.5e9-1.0e10) cm$^{-3}$ s for
Si and S in their solar abundance ratios.}
\end{deluxetable}

\begin{deluxetable}{lllllll}
\tabletypesize{\scriptsize} \tablecaption{Hard Continuum Knots: Spectral Fits} \tablewidth{0pt} \tablehead{
\colhead{Region} & \colhead{$\chi^2, \chi^2/dof$} & \colhead{$N_H$} & \colhead{FS: $kT$} & \colhead{$n_et$} & \colhead{RS: $kT$} & \colhead{$n_et$} \cr
\colhead{} & \colhead{} & \colhead{($10^{22}$ cm$^{-2}$)} & \colhead{(keV)} & \colhead{(cm$^{-3}$ s)} & \colhead{(keV)} & \colhead{(cm$^{-3}$ s)} 
}
\startdata
CKE & 226.9, 1.22 & 0.30 (0.28-0.31) & 3.3 (3.0-3.5) & 1.e8 ($<$1.3e8) & [0.9] & 5.3e9 (3.9-7.1e9) \\
CKW & 221.9, 1.64 & 1.13 (1.08-1.19) & 3.0 (2.6-3.3) & 3.0e9 (2.7-3.3e9) & [0.9] & 1.3e11 (7.4e10-2.7e11) \\
\enddata
\end{deluxetable}
\end{document}